\documentclass[%
 aip,
 reprint,
 amsmath,amssymb,
 citeautoscript
]{revtex4-1}

\usepackage{graphicx}
\usepackage{dcolumn}
\usepackage{bm}
\usepackage{hyperref}

\usepackage[usenames,dvipsnames]{xcolor}
\usepackage{float}
\usepackage{tabularx}
\usepackage{dsfont} 

\usepackage{multirow}
\usepackage{braket}
\usepackage{mathtools}
\usepackage[
    output-decimal-marker={.},
    separate-uncertainty=true,
    per-mode=fraction
]{siunitx}
\usepackage{physics}
\usepackage{subfigure}
\usepackage{tcolorbox}
\usepackage[absolute]{textpos}
\usepackage{soul}

\renewcommand{\vec}[1]{\textbf{#1}}

\newcommand{\subtext}[1]{_{\mathrm{#1}}}

\newcommand{\gvec}[1]{\bm{#1}}

\hypersetup{
    colorlinks=true,                        
    linkcolor=blue,                         
    citecolor=blue,                          
    filecolor=magenta,                      
    urlcolor=blue                           
} 

\begin{document}

\title{Temperature-transferable tight-binding model using a hybrid-orbital basis}

\author{Martin~Schwade}
 \affiliation{Physics Department, TUM School of Natural Sciences, Technical University of Munich, 85748 Garching, Germany}
\author{Maximilian~J.~Schilcher}%
 \affiliation{Physics Department, TUM School of Natural Sciences, Technical University of Munich, 85748 Garching, Germany}
 \author{Christian~Rever\'{o}n~Baecker}%
 \affiliation{Physics Department, TUM School of Natural Sciences, Technical University of Munich, 85748 Garching, Germany}
 \author{Manuel~Grumet}%
 \affiliation{Physics Department, TUM School of Natural Sciences, Technical University of Munich, 85748 Garching, Germany}
\author{David~A.~Egger}%
\email{david.egger@tum.de}
 \affiliation{Physics Department, TUM School of Natural Sciences, Technical University of Munich, 85748 Garching, Germany}

\date{\today}

\begin{abstract}
Finite-temperature calculations are relevant for rationalizing material properties yet they are computationally expensive because large system sizes or long simulation times are typically required. 
Circumventing the need for performing many explicit first-principles calculations, tight-binding and machine-learning models for the electronic structure emerged as promising alternatives, but transferability of such methods to elevated temperatures in a data-efficient way remains a great challenge. 
In this work, we suggest a tight-binding model for efficient and accurate calculations of temperature-dependent properties of semiconductors.
Our approach utilizes physics-informed modeling of the electronic structure in form of hybrid-orbital basis functions and numerically integrating atomic orbitals for the distance dependence of matrix elements.
We show that these design choices lead to a tight-binding model with a minimal amount of parameters which are straightforwardly optimized using density functional theory or alternative electronic-structure methods. 
Temperature-transferability of our model is tested by applying it to existing molecular-dynamics trajectories without explicitly fitting temperature-dependent data and comparison to density functional theory.
We utilize it together with machine-learning molecular dynamics and hybrid density functional theory for the prototypical semiconductor gallium arsenide. 
We find that including the effects of thermal expansion on the onsite terms of the tight-binding model is important in order to accurately describe electronic properties at elevated temperatures in comparison to experiment.
\end{abstract}

\maketitle

\section{INTRODUCTION}
    Computational investigations of the microscopic characteristics of functional materials are important for designing and discovering new compounds.
    Growing computational power and the development of new methods have enabled investigations of materials of ever-increasing complexity and size.
    Large-scale molecular dynamics (MD) simulations with high accuracy are now possible because of recent advances in combining machine learning (ML) with MD (ML-MD) \cite{behler_first_2017, butler_machine_2018, deringer_machine_2019, deringer_origins_2021, unke_machine_2021, fedik_extending_2022}.

    However, for semiconductors and their use in technological applications, it is crucial to characterize the electronic structure in such large-scale/long-time dynamical simulations as well.
    That is, even when accurate MD trajectories of such scenarios are available, one still faces the challenge of predicting their electronic structure.
    Such is particularly relevant in the context of capturing thermal characteristics of electronic states in structurally disordered materials and the electron-phonon interactions present in such systems.
    This is because structurally large and disordered quantum systems present great computational barriers to pertinent methods relevant in this context, such as density functional theory (DFT).  
    For this reason, implementations of linear-scaling DFT methods have been proposed \cite{kohn_density_1996a, soler_siesta_2002, nakata_large_2020, prentice_onetep_2020}, but combining self-consistency with speed and accuracy remains an on-going challenge despite recent advances \cite{dogan_solving_2023}.
    
    From a perspective of computational efficiency, parameterized methods such as tight binding (TB) have the advantage that the self-consistency loop is skipped. 
    Furthermore, due to the real-space nature of TB the electronic interactions may be neglected beyond a certain cut-off distance providing sparse Hamiltonian matrices.
    In this context, it is relevant that the TB method can capture the electronic structure of materials using only relatively few parameters, which can be obtained by fitting to any higher-level electronic-structure method.
    ML models can be powerful tools for such tasks as well \cite{lewis_learning_2021, kulik_roadmap_2022, grisafi_electronicstructure_2023}, but they can also be significantly more complex when learning properties such as the electron density.
    TB models may offer the advantage of requiring reduced amounts of training data, which is especially important for finite-temperature calculations of the electronic structure that require transferring the model to structurally distorted materials systems.
    
    Perhaps one of the most significant advances in the context of TB was the work by Slater and Koster (SK), providing a table of matrix elements that allowed efficient parameterization \cite{slater_simplified_1954}. 
    Based on their work, there have been numerous successful applications of the TB method to various materials \cite{kwon_transferable_1994, jancu_empirical_1998, papaconstantopoulos_handbook_2015a}.
    However, including thermal effects in a transferable approach still remained challenging, often involving temperature-dependent parameters or models for additional corrections in order to produce accurate thermal trends \cite{sawamura_nearestneighbor_2017, mishra_temperature_2023}.
    To overcome this, different strategies were suggested including non-orthogonal TB models \cite{menon_nonorthogonal_1997}, environment-dependent (or three-body) TB \cite{tang_environmentdependent_1996, garrity_fast_2023} or more recently the combination of TB with ML \cite{nakhaee_machine_2020, wang_machine_2021, zubatiuk_machine_2021, schattauer_machine_2022, gu_neural_2022} or density functional TB with ML \cite{kranz_generalized_2018, li_density_2018, fan_obtaining_2022}.
    These methods succeeded in increasing the transferability of the approach but often imply higher complexity and computational cost especially when more training data are required for optimizing the parameters.
    
    In this work, we present a TB framework that exhibits a high level of transferability to differently-sized systems and temperatures while requiring only a minimal amount of parameters.
    Most importantly, we show that the parameters of our model are, once optimized on small-scale structures, fully transferable across a wide temperature range (excluding the effect of thermal expansion) without requiring refitting or explicitly fitting temperature-dependent data.
    The approach is general and may be unified with some of the aforementioned techniques, such as ML, to further enhance its performance.
    In particular, our formalism is general and may be applied to different types of hybrid orbitals or even combined with atomic-orbital bases.

    The motivation for developing our model are calculations of electronic-structure properties in thermally disordered semiconductors that require large-scale simulations. 
    MD trajectories are routinely becoming available now through ML-MD and computing the electronic structure in such scenarios remains challenging and motivates method development.
    We build our method specifically for this important task, employing a hybrid-orbital basis set and an atomic-orbital-like distance dependence of matrix elements, with the goal of keeping the amount of parameters involved in the TB scheme minimal.
    To this end, a single fitting procedure that combines training data from non-distorted crystals and validation data from small-scale MD is used. 
    It is demonstrated that our TB method is transferable to larger systems with lower structural symmetry, which is key for capturing electronic-structure effects in disordered semiconductors at finite temperatures.
    We establish the accuracy, temperature-transferability and computational efficiency of the method by calculating pertinent thermal electronic-structure effects for the prototypical semiconductor gallium arsenide (GaAs).

\section{RESULTS}
    
    \subsection{Tight binding with a hybrid-orbital basis}\label{sec:hyb_orb_basis}
    
    The TB formalism is based on Bloch's theorem \cite{bloch_ueber_1929} and the linear combination of atomic orbitals (LCAO) method \cite{pauling_principles_1929}: 
    electrons are assumed to be tightly bound to their respective nucleus, which results in electronic orbitals that are strongly localized in real space around the atomic center. 
    Orbital overlaps decrease rapidly with the distance between atoms and interactions beyond a certain cut-off are typically neglected.
    
    We start with the TB Hamiltonian and its matrix elements
    \begin{equation}\label{eq:tbham}
        H_{ij}^{\vec{k}} = \mel{\psi_i^{\vec{k}}}{\hat{H}}{\psi_j^{\vec{k}}},
    \end{equation}
    where $\psi^{\vec{k}}_i$ is the $i$-th Bloch state. 
    In TB, we write it as Bloch sum over localized electronic orbitals, $\chi_i$, e.g., atomic orbitals (AOs) or Wannier functions, as follows:
    \begin{equation}\label{eq:tbwave}
        \ket{\psi^{\vec{k}}_i} = \frac{1}{\sqrt{N}}\sum_{\vec{R}} \mathrm{e}^{\mathrm{i}\vec{k}\cdot\vec{R}} \chi_i(\vec{r} - \gvec{\tau}_i - \vec{R}).
    \end{equation}
    Here, $i$ is a composite index that specifies the electronic orbital and respective atom, $\gvec{\tau}_i$ denotes its position within the unit cell and $\vec{R}$ denotes a lattice translation vector.
    Note that the TB wavefunction is generally not an energy eigenstate of the Hamiltonian. Also, while Eq.~\ref{eq:tbwave} is in principle exact, in practice it represents an approximation because the exact form of the orbitals, $\chi_i$, is usually not known and approximated via the TB basis.
    Furthermore, the $\psi$'s in Eq.~\ref{eq:tbwave} do not produce an orthonormal basis set, which can be taken into account through a L\"owdin orthogonalization \cite{lowdin_non_1950,slater_simplified_1954}.
    
    Our goal is to develop a transferable and accurate TB method requiring that the model must be sensitive to structural changes and applicable to different system sizes and temperatures. 
    The challenge lies in efficiently and accurately incorporating effects of structural changes without introducing a large amount of new parameters. 
    An ideal model for this task is one where the TB parameters are independent of system size.
    
    Our first design choice to achieve this desired property is the use of a hybrid-orbital basis set formed via an LCAO ansatz~\cite{pauling_nature_1931, slater_directed_1931}. 
    We find this has the important advantage that the hybrid orbitals can closely reflect the crystal symmetry of the system, e.g., by being oriented along the bonding axis of two atoms. 
    This aspect has been exploited recently to parameterize TB Hamiltonians in terms of hybrid orbitals based on Wannier functions \cite{hossain_hybrid_2021, hossain_selfenergy_2021}. 
    We also note that the idea of using hybrid orbitals is reminiscent of so-called polarized atomic orbitals introduced in Ref.~\citenum{lee_polarized_1997}.
    Additionally, a hybrid-orbital basis allows for combining different AO characters (\textit{s}-, \textit{p}- and \textit{d}-states) into one basis function without increasing basis-set size. 
    These features enable us to construct a basis of minimal size without compromising the added complexity of higher-order AOs that were found to be necessary for accurate band-structure predictions of semiconductors such as silicon and GaAs \cite{chadi_tightbinding_1975, richardson_electron_1986}.
    
    Specifically, we can expand a hybrid-orbital wavefunction as 
    \begin{equation}\label{eq:hyborb}
        \chi_i(\vec{r}-\vec{R}) = \frac{1}{A_i}\sum_{lm} f_{lm}(\gvec{\omega}_i)\ket{\phi_{lm}(\vec{r}-\vec{R})},
    \end{equation}
    where $A_i$ is a normalization constant, each $f_{lm}$ is a function of the orientation axis, $\gvec{\omega}_i$, associated with the respective $\chi_i$ that determines the contribution of each of the involved AOs, $\phi_{lm}(\vec{r}-\vec{R})$.
    The exact form of $f_{lm}$ depends on the choice of the underlying hybrid orbital, which we will discuss in Sec.~\ref{sec:tbparams} and Appendix \ref{app:tables}. 
    
    Let our original Cartesian reference frame be denoted by $xyz$. 
    As in the SK model \cite{slater_simplified_1954}, we express each integral between electronic orbitals in a reference frame, $x^\prime y^\prime z^\prime$, in which the $z$-axis lies along the vector connecting two atoms. 
    The transformation between $xyz$ and $x^\prime y^\prime z^\prime$ is described by a rotation matrix, $U$. 
    As the expansion of $\chi_i$ is only dependent on its axis of orientation, $\gvec{\omega}_i$, we obtain its representation in $x^\prime y^\prime z^\prime$ by applying the rotation matrix
    \begin{equation}
        \gvec{\omega}_i^r = U \cdot \gvec{\omega}_i,
    \end{equation}
    where $\gvec{\omega}_i^r$ is the axis of orientation of the $i$-th orbital in the rotated reference frame. 
    Rotation of a spherical harmonic can in general be expressed as a linear combination of spherical harmonics of the same $l$.
    Hence, Eq.~\ref{eq:hyborb} is also valid in the $x^\prime y^\prime z^\prime$ frame \cite{wigner_group_2013}.
    We can now evaluate the integral between two hybrids as in Eq.~\ref{eq:hyborb} but in the rotated frame
    \begin{align}\label{eq:hyborb_me}\nonumber
        \mel{\chi_i(\vec{r})}{\hat{H}}{\chi_j(\vec{r} - \vec{R})} = &\frac{1}{A_iA_j} \sum_{ll^\prime m}f^*_{lm}(\gvec{\omega}^r_i)f_{l^\prime m}(\gvec{\omega}^r_j) \times \\
        &\mel{\phi_{lm}(\vec{r})}{\hat{H}}{\phi_{l^\prime m}(\vec{r}-\vec{R})},
    \end{align}
    where we used the fact that integrals between two AOs of different $m$ vanish.
    
    Next, we parameterize the matrix elements between the AOs appearing in Eq.~\ref{eq:hyborb_me}. 
    Let $v$ be a composite index that runs over all non-zero terms in the sum over $ll^\prime m$, we summarize the terms that depend on the orientation of two hybrid orbitals in a tensor
    \begin{equation}\label{eq:angpref}
        C_{vij}^{\vec{R}} \vcentcolon = f^*_{lm}(\gvec{\omega}^r_i) f_{l^\prime m}(\gvec{\omega}^r_j).
    \end{equation}
    Using the two-center approximation \cite{slater_simplified_1954}, we can write each integral as a product between a scaling parameter and a function of the distance between the respective atoms, denoted as $\Delta r$.
    We therefore approximate each AO-like integral in Eq.~\ref{eq:hyborb_me} by
    \begin{equation}\label{eq:AOint}
        \mel{\phi_{lm}(\vec{r})}{\hat{H}}{\phi_{l^\prime m}(\vec{r}-\vec{R})} \approx V_{ll^\prime m}^0 V_{ll^\prime m}(\Delta r),
    \end{equation}
    where $V_{ll^\prime m}^0$ are adjustable parameters that we shall refer to as TB parameters in what follows (see Sec.~\ref{sec:tbparams}). $V_{ll^\prime m}(\Delta r)$ is a function that determines the scaling of how the orbital interactions, i.e., the coupling terms in the TB Hamiltonian, depend on $\Delta r$. The exact shape of $V_{ll^\prime m}(\Delta r)$ and the optimization of the TB parameters by fitting the model to DFT energy eigenvalues will be discussed in Secs.~\ref{sec:distdep} and \ref{sec:paramoptim}, respectively.
    
    A Hamiltonian matrix element can then be calculated in the hybrid-orbital basis by combining Eqs.~\ref{eq:tbham}, \ref{eq:angpref} and \ref{eq:AOint}:
    \begin{equation}\label{eq:tbham_fin}
        H_{ij}^{\vec{k}} = \frac{1}{A_i A_j}\sum_{\vec{R}} \mathrm{e}^{\mathrm{i}\vec{k}\cdot\vec{R}} \sum_{v} C_{vij}^{\vec{R}} V_{v}(\Delta r) V_{v}^0.
    \end{equation}

    \subsection{Basis set and TB parameters}\label{sec:tbparams}
\begin{figure}
        \centering
        \includegraphics[width=\linewidth]{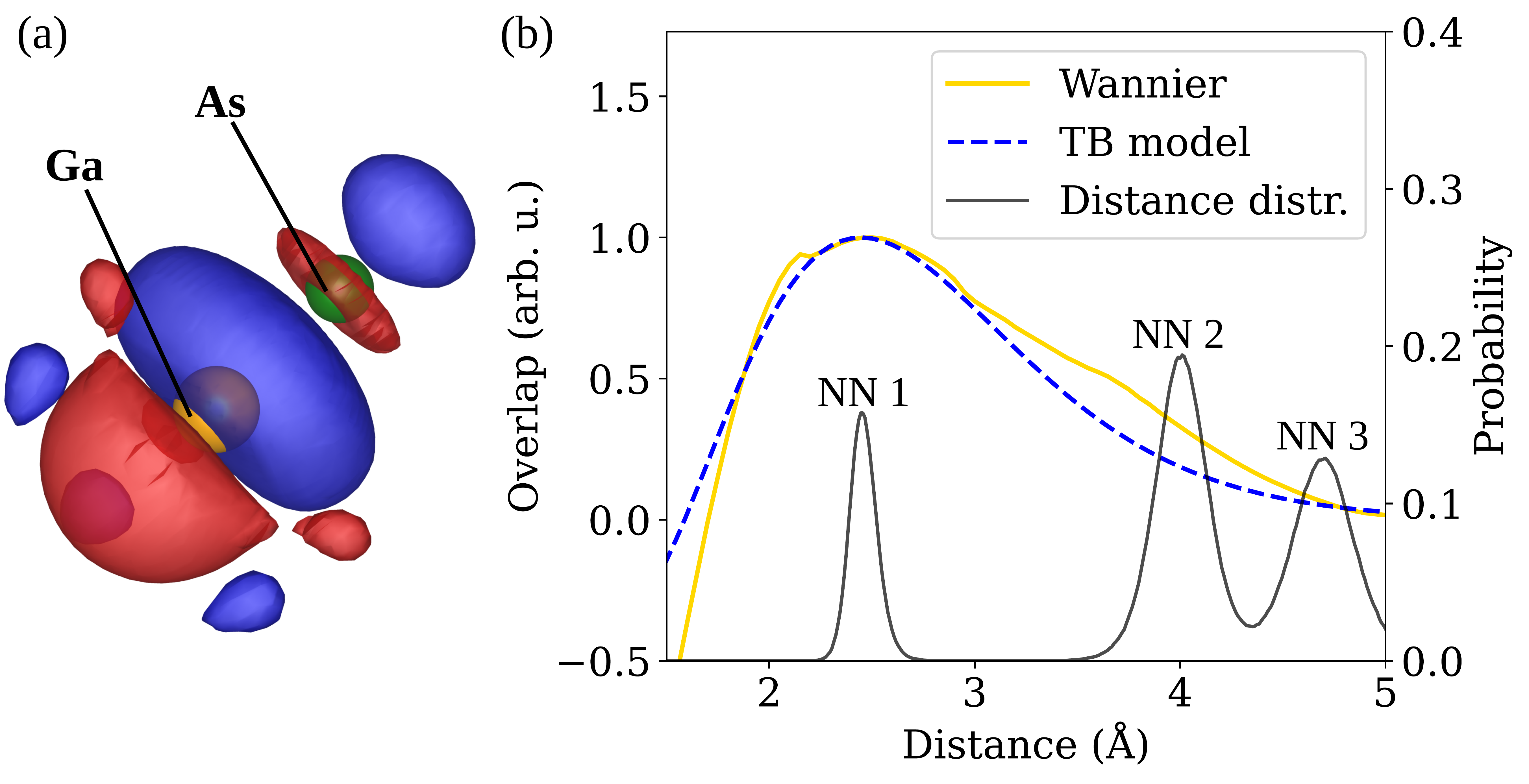}
        \caption{(a) Isosurface representation plotted with \texttt{VESTA} \cite{momma_vesta_2011} of the Wannier function (centered on Ga and pointing towards one of its As neighbors) obtained from projecting an $sp^3$ orbital onto the DFT wavefunctions. (b) Comparison of the distance dependence of the overlap computed in our model (dashed blue) and \texttt{Wannier90} (yellow). The distribution of interatomic distances as it occurred in the MD at \SI{400}{K} is given for reference where each peak corresponds to a nearest-neighbor (NN) distance.}
        \label{fig:wf_behav}
    \end{figure}

    We now discuss the choice of which hybrid orbitals to include in the basis set for a given material, the rotational behavior of Eq.~\ref{eq:angpref}, and the resulting TB parameters. 
    Our study focuses on tetrahedrally bonded semiconductors such as silicon, GaAs, or germanium as simple but relevant test systems. 
    We choose a hybrid orbital that combines an $sp^3$ hybrid and a rotated $d_{z^2}$ orbital, ensuring that the main symmetry axis of the $d_{z^2}$ orbital coincides with the axis of orientation of the respective hybrid orbital. Note that this rotated $d_{z^2}$ orbital is symmetric around the respective bond axis and does thus not break the crystal symmetry. Rather, it can be viewed as a hybrid orbital since it is constructed from a linear combination of all $d$ orbitals.
    This particular choice was motivated by the orbital character of the Wannier functions shown in Fig.~\ref{fig:wf_behav} (a): 
    the two lobes surrounding the Ga atom resemble an $sp^3$ orbital, while the ring and lobe structure centered on the As atom are reminiscent of  a rotated $d_{z^2}$ orbital. 
    Axes of orientation coincide with the vectors connecting the four nearest neighbors of a given atom in the non-distorted diamond crystal structure. 
    In practice, it is convenient to describe the axis of orientation, $\gvec{\omega}_i$, using its polar and azimuthal angle, $\theta_i$ and $\varphi_i$, respectively. 
    The angular prefactor in Eq.~\ref{eq:angpref} is then a function of these angles. 
    Following Ref.~\citenum{ivanic_rotation_1996}, we derived the explicit form of all expansion coefficients $f_{ll^\prime m}(\theta_i, \varphi_i)$ and provide them in Appendix \ref{app:tables}.
    
    Here, we note that the two-center approximation introduced in Sec.~\ref{sec:hyb_orb_basis} is unlikely to give reasonable results beyond nearest-neighbor integrals due to environmental effects of surrounding atoms \cite{garrity_fast_2023}.
    This issue can be addressed by choosing the cut-off radius to be the nearest-neighbor distance or, alternatively, by using a different parameter set for beyond-nearest-neighbor integrals. 
    In our model, we include beyond-nearest-neighbor interactions up to a cut-off radius (see Appendix \ref{app:tbfit}). In the case of GaAs, we naturally use different TB parameters to describe the interaction between Ga and As atoms (i.e., nearest neighbors) and Ga and Ga or As and As atoms (i.e., beyond nearest neighbors). The different environments associated with these interactions is thus captured in an effective manner through the TB parameters.
    We similarly introduce individual parameters for all onsite matrix elements, i.e., where $\Delta r = 0$. 
    A simplified approximation is chosen where onsite matrix elements are independent of the positions of surrounding atoms and thus also independent of temperature.
    
    Before proceeding, it is instructive to put the choice of the hybrid-orbital basis set of this work into the context of previous basis sets for tetrahedral semiconductors. 
    The $sp^3$ basis set was first extended by an $s^*$ orbital to improve the accuracy of the conduction bands \cite{vogl_semiempirical_1983}, thus increasing the Hamiltonian size to $10\times10$ with 13 parameters.
    It was then further extended by 5 $d$ orbitals \cite{jancu_empirical_1998}, increasing the Hamiltonian size to $20\times20$ with 27 parameters. Both models took into account nearest-neighbor interactions only. As described above, our model goes beyond nearest-neighbor interactions and contains 34 parameters for this reason.
    We expect our model to be more accurate than the one employing an $sp^3s^*$ basis and similarly or slightly less accurate than $sp^3d^5s^*$ while maintaining the minimal $8\times8$ Hamiltonian. This is achieved because extensions to the basis set are directly added onto all hybrid orbitals (see Eq.~\ref{eq:hyborb}) and do not change the dimensionality of the Hamiltonian matrix. However, it is for this reason that we cannot use the well-known SK relations for matrix elements as noted above, and rather employ the relations shown in Appendix \ref{app:tables}. On the other hand, the relatively small Hamiltonian in our method reduces the computational effort for diagonalization, which is the main bottleneck for larger systems.
    
    \subsection{Distance dependence of the matrix elements}\label{sec:distdep}
    
    Next to a hybrid-orbital basis the second main aspect of our TB method concerns the distance dependence of the matrix elements given by $V_{v}(\Delta r)$ in Eq.~\ref{eq:tbham_fin}.
    Note that, as discussed above, this distance-dependence function only affects coupling matrix elements and not onsite matrix elements.
    A common ansatz is to use a polynomial times exponential function for each TB parameter \cite{kwon_transferable_1994, papaconstantopoulos_slater_2003}. 
    As an example, we consider using 3 parameters per polynomial and 1 parameter for the exponential function \cite{papaconstantopoulos_slater_2003}.  
    In the case of GaAs and a TB basis with $s$, $p$, and $d$ orbitals, the total amount of resulting parameters just for modeling the distance dependence would then be 120.
    While flexible, the large number of parameters renders the parameter optimization challenging.
    Because our aim is to keep the number of adjustable parameters minimal we follow a different strategy.
    
    As described in Eq.~\ref{eq:hyborb_me}, the hybrid basis set required computation of AO-like integrals.
    For each AO, we assume the same, \textit{s}-type radial part when evaluating the integrals. 
    Each radial part is determined by only two parameters per atomic species present in the system, which are the principal quantum number, $n$, and the effective core charge, $Z\subtext{eff}$. 
    The radial part is then an \textit{s}-orbital of given $n$ with scaling parameter, $Z\subtext{eff}$.
    In this way, the number of parameters for modeling the distance dependence only depends on the number of atomic species present in the material and is independent of the number of interactions, which is favorable for calculating the properties of large and disordered systems. For GaAs, this results in a total of 4 parameters for modeling the distance dependence.
    
    To parameterize the distance-dependence functions in Eq.~\ref{eq:tbham_fin}, we mimic the distance-dependence of Wannier functions: 
    Fig.~\ref{fig:wf_behav} (b) shows the value of the matrix element where both orbitals are oriented along the bonding axis when one atom is shifted in the unit cell. 
    To obtain these data, we run a DFT calculation with \texttt{VASP} \cite{kresse_efficient_1996} for each of these structures and then compute Wannier functions using \texttt{Wannier90}~\cite{mostofi_updated_2014} (see Appendix \ref{app:dft} for details).
    The maximum of the curve occurs at the equilibrium distance, which corresponds to the peak labeled ``NN1'' in Fig.~\ref{fig:wf_behav} (b).
    We set $n$ to be equal for Ga and As since they are found in the same period of the periodic table. 
    Next, we adjust the values $Z\subtext{eff}$ until the maximum of the curve computed from the integrals between the respective hybrid orbitals coincides with the nearest-neighbor distance.  
    For $n=2, 3, 4$ we find that the resulting fits (see Sec.~\ref{sec:paramoptim}) exhibit the smallest validation error when $n=3$, i.e., the errors are $n=2$: \SI{0.102}{eV}, $n=3$: \SI{0.098}{eV}, $n=4$: \SI{0.128}{eV}. 
    Note that $n=2$ would not be a reasonable choice in view of the basis set we are employing because $d$ orbitals do not exist for that principal quantum number.
    Therefore, $n=3$ is used for which we find $Z\subtext{eff, Ga} = 9$ and $Z\subtext{eff, As} = 13$. 
    Since our model includes beyond nearest-neighbor interactions as described above, the remaining deviations occurring beyond nearest-neighbor distances (see Fig.~\ref{fig:wf_behav} (b)) can still be captured when optimizing the TB parameters.
    We then determine the distance-dependent scaling factor, $V_{ll^\prime m}(\Delta r)$ in Eq.~\ref{eq:AOint}, from the value of the integral between the respective AOs at a given $\Delta r$
    \begin{equation}\label{eq:distdep}
        V_{ll^\prime m}(\Delta r) = \braket{\phi_{lm}(\vec{r})}{\phi_{l^\prime, m}(\vec{r} - \Delta r\hat{\vec{e}}_z)}.
    \end{equation}
    
    Using this scheme to construct the full Hamiltonian matrix in Eq.~\ref{eq:tbham_fin} would require evaluating many 3D integrals, which is computationally demanding. 
    Hence, we interpolate the integrals in Eq.~\ref{eq:distdep} as a function of distance using cubic splines, which significantly speeds up the procedure computationally. 
    Inspired by the above findings, the values of $n$ and $Z\subtext{eff}$ are chosen such that the maximum of the distance-dependent function coincides with the nearest neighbor distance in the material and the rate of decay is similar to that observed for the respective Wannier functions (see Fig~\ref{fig:wf_behav}).
    Hence, the two parameters $n$ and $Z\subtext{eff}$ are chosen based on physical considerations and not optimized further.
    This is helpful to imbue the model with sufficient transferability while maintaining important physical characteristics of the material within the approach.
    
    \subsection{Workflow of our tight binding approach}\label{sec:paramoptim}
    \begin{figure}
        \centering
        \includegraphics[width=\linewidth]{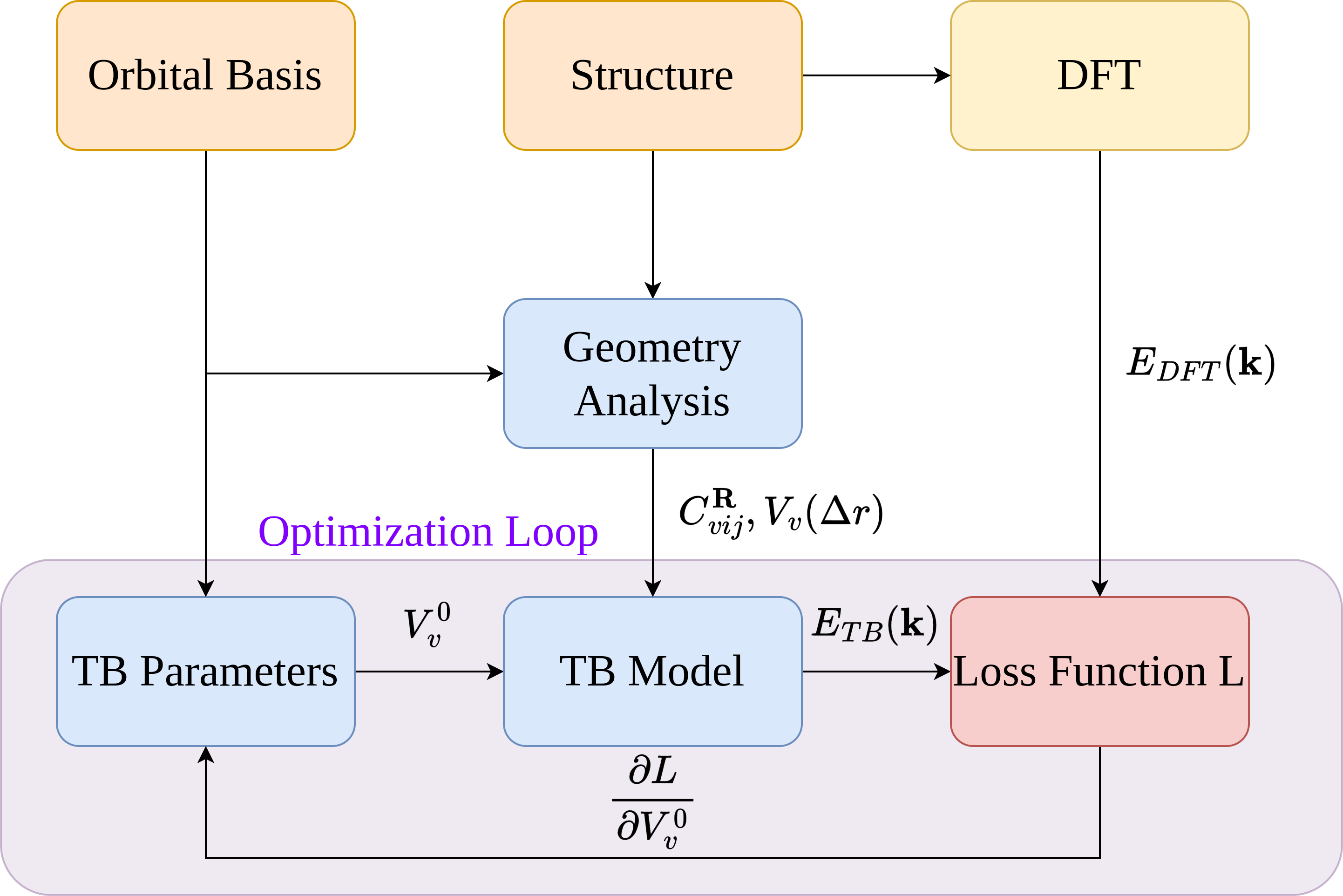}
        \caption{Workflow of our TB method to optimize the parameters, which are determined by minimizing the error between TB (blue) and DFT (yellow).}
        \label{fig:tbfit}
    \end{figure}
    Integration of the hybrid-orbital basis set and distance-dependent modeling of our TB model into an efficient workflow is important. We implemented our TB method in a standalone code library written in the Julia programming language.
    Our TB workflow is depicted in Fig.~\ref{fig:tbfit}: 
    it starts with information on the structure of the material (unit cell and lattice vectors) and orbital basis to be used in the TB scheme. 
    A geometry analysis yields angular and distance-dependent factors of Eq.~\ref{eq:tbham_fin} in a preprocessing step, since they have to be calculated only once for each material. 
    Since our TB basis set is not dependent on the structure, i.e., it does not depend on cell size or lattice vectors, we can transfer the TB basis to any system size after it has been parameterized once. 
    Finally, when the TB Hamiltonian is calculated its resulting energy eigenvalues are compared to the corresponding data calculated in DFT, and TB parameters are updated depending on the mismatch, calculated from the loss function (see Appendix \ref{app:tbfit} for details). 
    The procedure is repeated until satisfactory agreement is reached.
    
    In the iterative procedure of optimizing the TB parameters we combine information of two types of data sets, a training and a validation data set, as is also customary in the context of ML \cite{murphy_probabilistic_2022}. 
    The training data are DFT eigenvalues of the primitive unit cell obtained in electronic bandstructure calculations and are used to optimize the TB parameters. 
    The validation data are DFT eigenvalues of distorted structures obtained in small-scale MD calculations and are used to prevent over-fitting via early stopping. 
    More details about the training and validation data sets and the optimization procedure are found in Appendix \ref{app:dft} and \ref{app:tbfit}, respectively.
    \begin{figure}
        \centering
        \includegraphics[width=\linewidth]{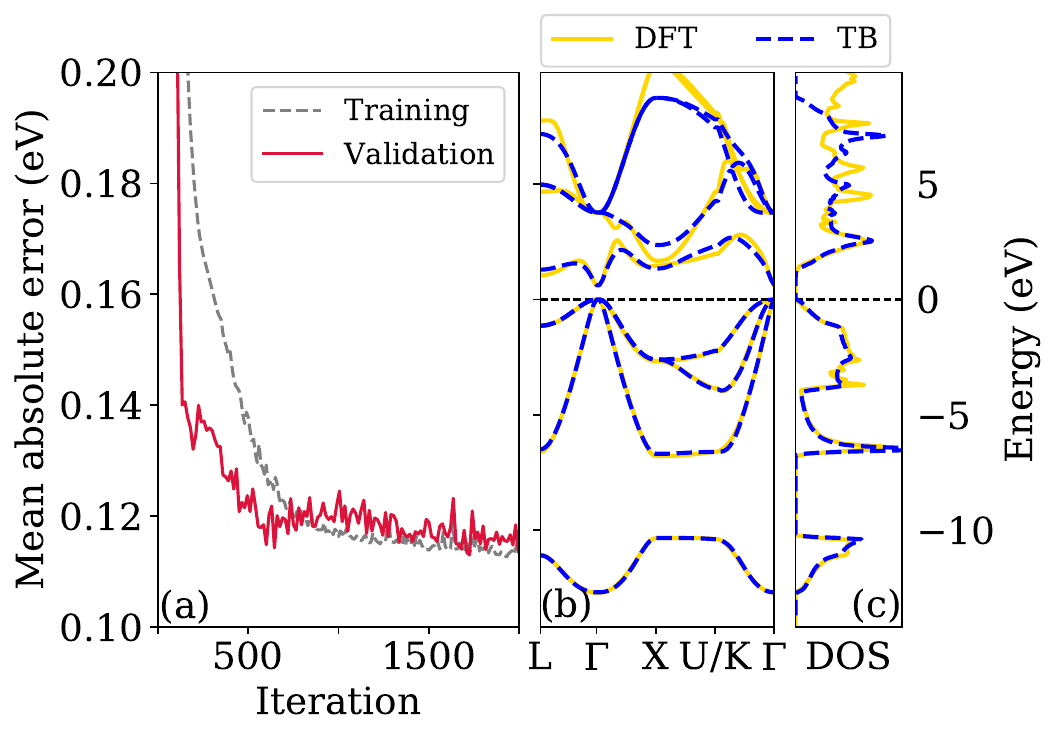}
        \caption{(a) Training (grey) and validation error (red) during parameter optimization, quantified by the mean-absolute error (see Appendix \ref{app:effbands}). Electronic bandstructure (b) and density of states (DOS) (c) of GaAs computed by TB (dashed blue) and DFT (yellow). Zero on the $y$-axis is set to the valence-band maximum.}
        \label{fig:TBDFTbands}
    \end{figure}%
    
    We apply the TB model obtained in this way to a static unit cell of GaAs (see Appendix \ref{app:dft} for details).
    Fig.~\ref{fig:TBDFTbands} (a) shows the iterative evolution of training and validation errors, illustrating convergence of the optimization procedure. Note that the training and validation errors are calculated for different numbers of bands, $\vec{k}$-points, and weights and can therefore not be compared quantitatively.
    The TB model is then used to compute the electronic bandstructure and density of states (DOS) of GaAs compared to DFT results (see Figs.~\ref{fig:TBDFTbands} (b) \& (c)). 
    The agreement of TB with DFT is overall reasonable as quantified by comparing further electronic-structure observables: 
    the bandgaps are found to be $E^{\mathrm{TB}}\subtext{gap} = \SI{0.63}{eV}$ and $E^{\mathrm{DFT}}\subtext{gap} = \SI{0.60}{eV}$, showing that the deviation is small compared to the value itself.
    We calculate effective masses from TB that are reasonably close to DFT with a mismatch of ${<}0.01\,m\subtext{e}$ for electron and light-hole mass and $0.04\,m\subtext{e}$ for heavy-hole mass.
    It is worth noting that the agreement of the valence bands is somewhat better than for the conduction bands. 
    This reflects a general limitation of TB that is well-known: localized atomic orbitals employed in TB are not an optimal representation for the delocalized conduction band states in GaAs. 
    Not taking into account the upper two conduction bands, the mean-absolute error (MAE) between the DFT and TB bandstructure is calculated to be \SI{0.10}{eV} (see Appendix \ref{app:effbands} for details).
    We consider this level of agreement to be satisfactory for the main envisioned purpose of our TB method, which are calculations of the thermal electronic properties of semiconductors, since it is smaller than the bandgap of typical semiconductors such as GaAs.

    \subsection{Finite-temperature electronic structure}\label{sec:gaas_tempdep}
        
    \begin{figure}
        \centering
        \includegraphics[width=\linewidth]{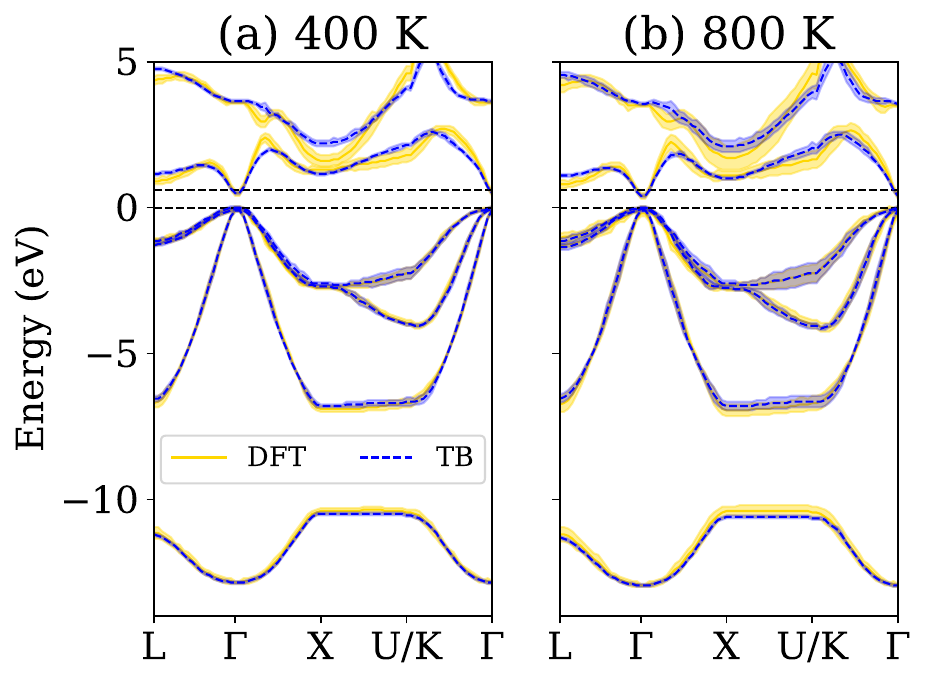}
        \caption{Effective electronic bandstructure of GaAs along high-symmetry paths at \SI{400}{K} (a) and \SI{800}{K} (b) calculated using our TB model (dashed blue) and DFT (yellow).}
        \label{fig:tdep_bands_gaas}
    \end{figure}
    We investigate the performance of our TB method for finite-temperature electronic-structure calculations. 
    To this end, we apply the model parameterized as described above to snapshots of DFT-MD trajectories at different temperatures in order to study transferability of it to large and disordered structures.
    Fig.~\ref{fig:tdep_bands_gaas} shows effective bandstructures of GaAs computed at 400 and \SI{800}{K} with our TB method and DFT (see Appendix \ref{app:effbands} for details). 
    Effective bandstructures calculated with TB and DFT show a high level of agreement: 
    comparing the individual mean values across all bands (again not taking into account the upper two conduction bands), we receive a MAE of \SI{0.11}{eV} and \SI{0.12}{eV} for \SI{400}{K} and \SI{800}{K}, respectively.
    This finding shows that our TB method accurately captures thermal trends of the electronic properties since the error between TB and DFT barely depends on temperature.
    As a case in point, both TB and DFT find that the bandgap decreases with temperature and that certain degeneracies are lifted as a consequence of dynamic symmetry breaking.
    \begin{figure}[b]
        \centering
        \includegraphics[width=\linewidth]{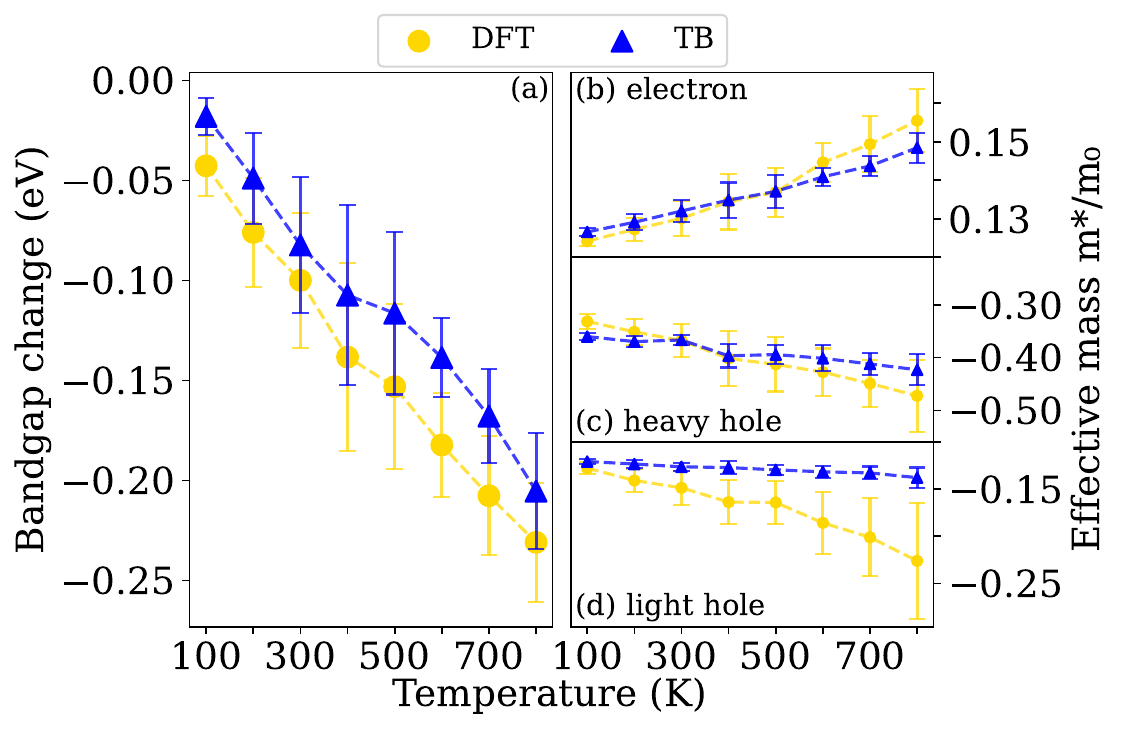}
        \caption{(a) Comparison of the change of the bandgap with temperature for GaAs calculated by TB (blue triangles) and DFT (yellow circles). The bandgap change is calculated with respect to the value obtained at \SI{0}{K} (TB: \SI{0.63}{eV}, DFT: \SI{0.60}{eV}). Corresponding temperature dependence of light-hole (b), heavy-hole (c) and electron effective mass (d).}
        \label{fig:bandgap_trend}
    \end{figure}
    
    The DFT calculations intrinsically account for changes in the electronic structure that are due to thermally-induced structural dynamics because the ground-state of each MD snapshot is calculated self-consistently. 
    TB does not self-consistently calculate the ground state and it is instructive to analyze how structural changes still enter the approach by revisiting Eq.~\ref{eq:tbham_fin}. First, in general any change of the structure can cause changes in the relative orientations between two hybrid orbitals $i$ and $j$, despite the orientation axis of $i$ and $j$ being generally maintained. 
    These changes are captured via the angular-dependent factor, $C_{ijv}^{\vec{R}}$, in Eq.~\ref{eq:tbham_fin}. 
    Second, along the MD trajectory the atomic distances increase or decrease, which in TB is accounted for by the distance-dependent factor, $V_v(\Delta r)$. 
    These quantities are evaluated for each of the instantaneous configurations occurring in an MD trajectory by the “Geometry Analysis” module of our workflow (see Fig.~\ref{fig:tbfit}).
    By design of our TB approach, these adaptions do not imply any update to the TB parameters, $V_{ll^\prime m}^0$.
    
    The calculated finite-temperature electronic structure provides access to important thermal characteristics of semiconductors.
    First, we compute the temperature-dependence of the bandgap of GaAs with our TB method and DFT (see Fig.~\ref{fig:bandgap_trend}). 
    Remarkably, both methods yield a very similar thermal trend of the bandgap, with the TB method offering an accuracy that is similar to DFT at a much lower computational cost (see below for further results and analysis).
    Another relevant finite-temperature electronic property is the temperature-dependence of the effective hole and electron masses. 
    Calculating effective masses at different temperatures proves challenging in practice, see Appendix \ref{app:effbands} for details.
    Most crucially, the trends for the electron and heavy-hole effective mass from DFT and TB agree well while, by contrast, the light-hole mass exhibits a large discrepancy. 
    The temperature-induced changes of electron and heavy-hole masses are captured well in TB because they are due to dynamic changes of the coupling matrix elements in the TB Hamiltonian.
    However, the TB valence band at $\Gamma$ that is associated with the light-hole effective mass mainly depends on onsite matrix elements of the TB Hamiltonian. 
    Because the onsite matrix elements are temperature-independent in our model as we discussed above, the TB light-hole effective mass remains constant as temperature increases and does not capture well the trend obtained in DFT.
    This points out a possible route for improving our model in future work in order to go beyond the approximation that onsite matrix elements are independent of the atomic positions.
    
    \subsection{ML-MD and hybrid density functional theory}\label{sec:mlmd}
    
    So far we have benchmarked our TB method on the basis of DFT calculations.
    Two aspects of our TB implementation will now be investigated that we think are crucial when using the model for truly predictive finite-temperature electronic-structure calculations of semiconductors:
    first, it will be assessed whether TB can be combined with ML-MD in order to access large system sizes and long time scales. 
    Second, we will examine to what extent the TB model can be parameterized by more accurate electronic-structure calculations in order to provide a more reliable quantitative assessment on the thermal electronic behavior of semiconductors.
    
    Hence, we first combine TB with an ML-MD method as implemented in \texttt{VASP} \cite{jinnouchi_phase_2019, jinnouchi_onthefly_2019, jinnouchi_descriptors_2020}. 
    We showed above that TB allows for calculating temperature-dependent electronic properties with DFT-like accuracy at a significantly lower computational cost. 
    ML-MD allows for generating finite-temperature MD trajectories with comparable accuracy to traditional DFT-MD in a computationally more efficient manner. 
    Combining the two methods promises computationally efficient electronic-structure calculations of disordered materials at finite temperatures.
    
    We apply our TB method to MD trajectories of GaAs that were computed with ML-MD and DFT in order to obtain the temperature-dependence of the bandgap (see Appendix \ref{app:dft}).
    First of all, Fig.~\ref{fig:bg_vs_T_mlmd} shows that temperature-dependent bandgaps computed with DFT on either ML-MD or DFT-MD trajectories gives virtually the same result, with a MAE of $<$\SI{0.01}{eV} for the bandgaps.
    Therefore, DFT-MD can essentially be replaced by ML-MD for the task of providing temperature-dependent trajectories in order to compute bandgaps. 
    More importantly in the context of this work, Fig.~\ref{fig:bg_vs_T_mlmd} (a) shows that the TB data agree well with the full DFT-MD result, with a MAE of \SI{0.026}{eV} for TB+ML-MD which is reasonably small and  similar to the value for TB+DFT-MD of \SI{0.031}{eV}.
    
    The findings are encouraging because they show that our TB model works equally well when combined with either MD method (\textit{cf.}~Fig.~\ref{fig:bandgap_trend} and Fig.~\ref{fig:bg_vs_T_mlmd}).
    Assessing the total computing time of the DFT-MD run, we estimate that for our specific case the usage of ML-MD shortened the average run-time by a factor of 40, which may be reduced by further optimizing the ML-MD procedures. 
    Comparing the run-time of the entire unfolding workflow of our TB implementation and DFT using the PBE functional, we observe that using TB leads to a speed-up by a factor of around 500 compared to DFT (excluding the one-time cost of parameter optimization). 
    Taken together, combining ML-MD and TB provides a powerful tool to accurately compute temperature-dependent electronic structures of large and disordered systems at longer time-scales.
    In this context, it is worth noting that the algorithm employed in this work scales approximately to the third power with the number of atoms due to full matrix diagonalization as is the case also with many typical DFT implementations.
    However, the method can scale linearly with the number of atoms once its inherent locality is exploited (see, e.g., Ref.~\cite{goedecker_linear_1999}).
    
    \begin{figure}
        \centering
        \includegraphics[width=\linewidth]{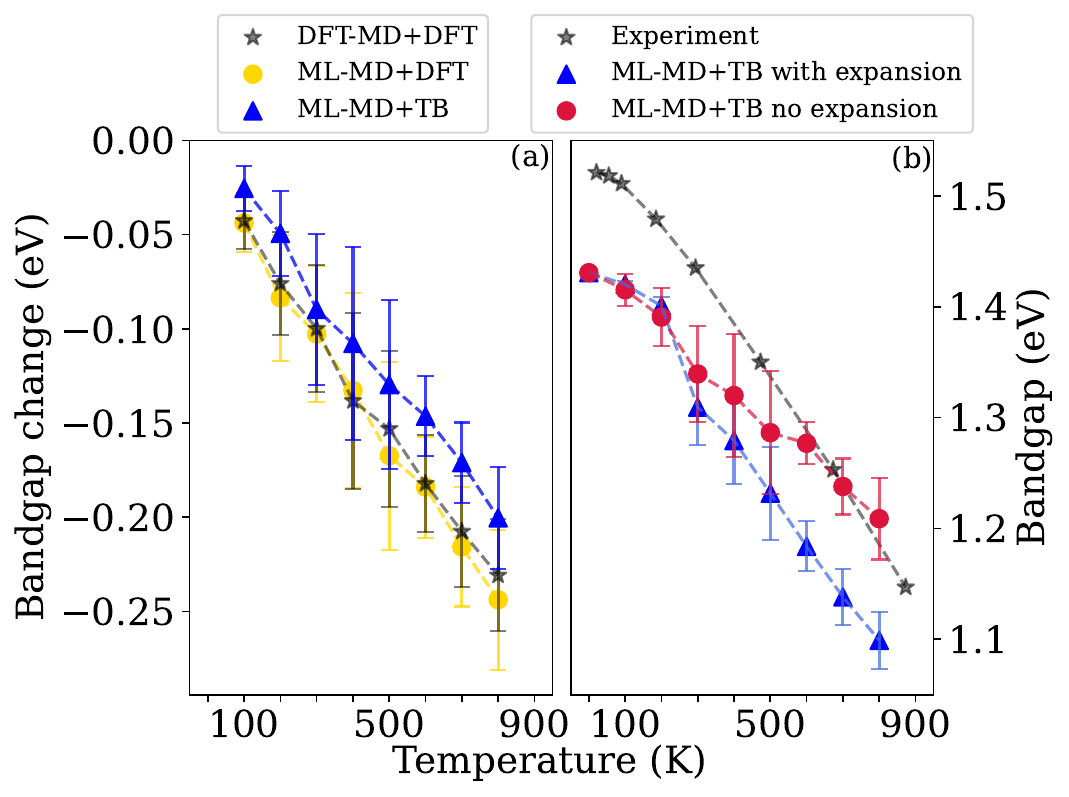}
        \caption{(a) Comparison of the temperature-dependent bandgap change obtained with DFT for DFT-MD trajectories, (black stars), DFT for ML-MD trajectories (yellow circles), and TB for ML-MD trajectories (blue triangles). The bandgap change is calculated with respect to the value obtained at \SI{0}{K}. (b) Comparison of the absolute bandgap value computed by TB fitted to HSE data, including (blue triangles) and excluding (red circles) thermal expansion effects, to experimental data \cite{panish_temperature_1969, sturge_optical_1962} (black stars). }
        \label{fig:bg_vs_T_mlmd}
    \end{figure}
    
    Furthermore, our formulation of TB in principle allows any level of electronic-structure theory for parameterizing the involved TB parameters, similar to previous strategies applied in the context of parameterizing electronic couplings of molecules \cite{manna_prediction_2018a}.
    This is important for obtaining reliable finite-temperature electronic properties and other quantities, e.g., electron-phonon couplings, because the accuracy of TB can be controlled by using more sophisticated electronic-structure theory in its parameterization.
    To illustrate the premise of this particular aspect of the TB method, we investigate the temperature dependence of the bandgap using TB parameters fitted to hybrid density functional theory. 
    Specifically, we employ the HSE exchange-correlation functional \cite{heyd_hybrid_2003, krukau_influence_2006} because for semiconductors such as GaAs this functional is well-known to provide electronic-structure properties that are more accurate than those calculated by PBE \cite{paier_screened_2006, marsman_hybrid_2008}.
    We investigate whether the known improvement of HSE compared to PBE data on the DFT level will be reflected in the TB results as well.
    
    In Fig.~\ref{fig:bg_vs_T_mlmd} (b), we show the temperature-dependence of the absolute bandgap value computed by TB in comparison to experimental bandgap data for GaAs \cite{panish_temperature_1969, sturge_optical_1962}. 
    When comparing to experimental data we included the effect of zero-point renormalization for the \SI{0}{K} bandgap data through the one-shot method \cite{zacharias_oneshot_2016} as implemented in \texttt{VASP}.
    While the model as parameterized above seems to capture the experimental trend well up to approximately room temperature it progressively deviates when temperature is raised further. 
    We therefore implement a simplified, semi-empirical scheme that accounts for the effect of thermal expansion on the onsite parameters of the TB model.
    To this end, we conducted additional ML-MD runs at lattice constants that increase linearly with increasing temperature according to the experimental expansion coefficient \SI{6.0e-6}{K^{-1}} \cite{yu_fundamentals_2010}, and fitted a new set of TB parameters for each lattice constant (see App.~\ref{app:tbfit}). 
    Through this ad hoc procedure, we find that including the effect of thermal expansion in TB drastically improves the calculation of the temperature-dependent bandgap of GaAs and results in excellent agreement with the experimental data. This assessment is based on the observation that the remaining ${\approx}$\SI{0.1}{eV} error compared to the absolute experimental data is essentially independent of temperature and already present in \SI{0}{K} HSE-DFT calculations.
     Furthermore, we would like to mention that the TB model is by a factor of around 40000 faster than HSE-DFT for the primitive-cell calculation at \SI{0}{K}, because the efficiency of TB does not depend on the underlying exchange-correlation functional used in DFT. 
     It is also noted that tests of the thermal expansion procedure for DFT and TB using the PBE functional find that the MAEs reported above (\textit{cf.}~Fig.~\ref{fig:bg_vs_T_mlmd} (a)) only change by 0.01-0.02\, eV. 
     This shows that the thermal-expansion procedure introduces a systematic error for our TB model that is small compared to the absolute value of the band gap as well as to its change with temperature.
    We hypothesize that using more accurate electronic-structure methods in the parameterization, such as optimally-tuned range-separated hybrid functionals \cite{wing_comparing_2019}, may further improve the quantitative accuracy of TB.

     Finally, we summarize the limitations of our model and provide directions for future developments. 
     First, the idea of using a hybrid-orbital basis and the formalism we have provided can in principle be applied to a wide class of semiconducting materials. Relevant in this context, the choice of hybrid orbital (defined by a particular linear combination of atomic orbitals) depends on the bonding environment of the respective atom, as shown in Fig.~1 in Ref.~\citenum{hossain_hybrid_2021}.
     However, it should be noted that the basis set we discussed here was chosen for GaAs in particular, and more specifically the zincblende or diamond structure and a tetrahedral bonding environment. 
     Adaptation to different phases of GaAs or other materials more broadly is expected to be possible because the presented formalism of hybrid orbitals is general and can be applied to different types of basis functions, including atomic orbitals. 
     Furthermore, while GaAs shows significant atomic displacement at 800\,K we expect that the model will be less accurate for materials with very large atomic displacements, because atomic environments may vary significantly from the training set.  
     Here, we believe that including three-center interactions as in, e.g., Ref.~\citenum{garrity_fast_2023}, which currently are neglected in our approach, as well as treating onsite matrix elements as being explicitly dependent on the positions of atomic neighbors,\cite{tan_transferable_2016a} are possible routes for further improvements. 
     Last but far from least, the current tight-binding model does not take spin-orbit coupling into account, which can be an important effect depending on the system and its atomic composition. For hybrid orbitals, this may imply using the well-known spin-orbit coupling matrices for atomic orbitals (see, e.g., Ref.~\citenum{gu_computational_2023}) and derive their representation in a basis of hybrid orbitals.
    
\section{CONCLUSION}
    In summary, we developed and implemented a TB model for temperature-dependent electronic properties of semiconductors with an accuracy similar to DFT but at much lower computational cost. 
    We achieved this by extending the SK approach and two-center approximation to a hybrid-orbital basis set and employing an AO-like distance dependence for overlap integrals. 
    This strategy led to a small number of parameters that we fitted to \SI{0}{K} DFT data. 
    Temperature-transferability of our method was demonstrated by accurate bandstructures, bandgaps and effective masses at finite-temperature compared to DFT for the prototypical semiconductor GaAs.
    We also found that the approximation of temperature-independent onsite parameters is reasonable for GaAs and discussed limitations of it regarding light-hole effective masses of this material. 
    The extensions of the TB framework presented here may support further developments of related models for various materials, enabling accurate computations of thermal electronic properties in large-scale, disordered systems.
    Such computational scenarios are becoming increasingly available now through ML-MD. 
    In conclusion, the present TB approach may hold many practical applications that promise to aid the discovery of novel functional materials through deeper understanding of their dynamic electronic properties. 
    For example, application of the method to other III-V semiconductors, alloys, or doped systems is anticipated to be relatively straightforward. 
    Applications to semiconducting materials more generally motivate further developments, e.g., regarding the influence of spin-orbit coupling, three-center interactions, and extending the basis set to different hybrid and atomic orbitals.
    
\section*{ACKNOWLEDGEMENTS}
Funding provided by the Alexander von Humboldt-Foundation in the framework of the Sofja Kovalevskaja Award, endowed by the German Federal Ministry of Education and Research, by TUM.solar in the context of the Bavarian Collaborative Research Project Solar Technologies Go Hybrid
(SolTech), and by TU Munich - IAS, funded by the German Excellence Initiative and the European Union
Seventh Framework Programme under Grant Agreement No.\ 291763, are gratefully acknowledged.
The authors further acknowledge the Gauss Centre for Supercomputing e.V.\ for funding this project by providing computing time through the John von Neumann Institute for Computing on the GCS Supercomputer JUWELS at Jülich Supercomputing Centre.

\section*{DATA AVAILABILITY STATEMENT}
    The data that support the findings of this study are openly available in Ref.~\citenum{schwade_zenodo_2023}.
    \appendix

    \section{Further technical details}\label{app:tables}
    \begin{table}[H]
        \caption{Rotational representation of the \textit{spd} hybrid-orbital components.}
        \begin{ruledtabular}
            \renewcommand{\arraystretch}{1.3}
            \begin{tabular}{l l}
                Orbital & $f_{lm}(\theta, \varphi)$ \\ \colrule
                $s$ & 1 \\
                $p_x$ & $\sin(\theta)\cos(\varphi)$ \\
                $p_y$ & $\sin(\theta)\sin(\varphi)$ \\
                $p_z$ & $\cos(\theta)$ \\
                $d_{yz}$ & $\sqrt{3}\sin(\theta)\cos(\theta)\sin(\varphi)$ \\
                $d_{xz}$ & $\sqrt{3}\sin(\theta)\cos(\theta)\cos(\varphi)$\\
                $d_{xy}$ & $\sqrt{3}\sin^2(\theta)\sin(\varphi)\cos(\varphi)$ \\
                $d_{x^2-y^2}$ & $\frac{\sqrt{3}}{2} \left(\sin^2(\theta)\cos^2(\varphi) - \sin^2(\theta)\sin^2(\varphi)\right)$\\
                $d_{z^2}$ & $ \frac{1}{2}\left(3\cos^2(\theta)-1\right)$\\
            \end{tabular}
        
        \end{ruledtabular}
        \label{tab:exp_coef_hyb}
    \end{table}
    
    \begin{table}[H]
        \caption{Orientation-dependent prefactors of all ten TB parameters.}
        \begin{ruledtabular}
            \renewcommand{\arraystretch}{1.3}
            \begin{tabular}{l l}
                TB Matrix Element & Contributions \\ \colrule
                $V_{ss\sigma}$ & $f_{s, 1}f_{s, 2}$ \\
                $V_{sp\sigma}$ & $f_{s, 1}f_{pz, 2} - f_{pz, 1}f_{s, 2}$ \\
                $V_{pp\sigma}$ & $f_{pz, 1}f_{pz, 2}$ \\
                $V_{pp\pi}$ & $f_{px, 1}f_{px, 2}$ + $f_{py, 1}f_{py, 2}$ \\
                $V_{sd\sigma}$ & $f_{s, 1}f_{dz2, 2} + f_{dz2, 1}f_{s, 2}$ \\
                $V_{pd\sigma}$ & $f_{pz, 1}f_{dz2, 2} - f_{pz, 1}f_{dz2, 2}$ \\
                \multirow{2}{*}{$V_{pd\pi}$} & $f_{px, 1}f_{dxz, 2} - f_{dxz, 1}f_{px, 2}$ \\ & $+ f_{py, 1}f_{dyz, 2} - f_{dyz, 1}f_{py, 2}$ \\
                $V_{dd\sigma}$ & $f_{dz2, 1}f_{dz2, 2}$ \\
                $V_{dd\pi}$ & $f_{dxz, 1}f_{dxz, 2} + f_{dyz, 1}f_{dyz, 2}$ \\
                $V_{dd\delta}$ & $f_{xy, 1}f_{xy, 2}$ \\
            \end{tabular}
        \end{ruledtabular}
        \label{tab:my_label}
    \end{table}

    \section{Details of DFT and MD calculations}\label{app:dft}
    DFT calculations were performed with \texttt{VASP} \cite{kresse_efficient_1996} and the PBE functional \cite{perdew_generalized_1996} unless otherwise stated, using a plane-wave cut-off energy of \SI{250}{eV}.
    To calculate the Wannier functions from DFT by \texttt{Wannier90} \cite{mostofi_updated_2014}, we use a $5\times5\times5$ \vec{k}-point grid, define 8 projections with $sp^3$ orbitals, and set the disentanglement window to $[\SI{-13}{eV}, \SI{15}{eV}]$. The disentanglement procedure is converged while the number of iterations to minimize the spread is set to 0 to conserve the shape of the Wannier functions.
    For training data, the energy eigenvalues discretized on an $11\times11\times11$ Monkhorst-Pack grid of the first Brillouin zone for the 2-atom primitive cell (lattice constant: \SI{5.65}{\text{\AA}}) were used, including only the 56 $\vec{k}$-points of the irreducible part. 
    The small-scale MD calculations for producing the validation data considered a cubic 8-atom cubic cell ($5\times5\times5$ \vec{k}-point grid) at a temperature of \SI{300}{K}. The validation set comprises a total of 22 electronic bands in order to simplify the fitting procedure for the supercell calculations.
    Validation errors were calculated by averaging over 10 random snapshots of this MD. 
    Temperature-dependent electronic structure calculations were performed for a 128 atom non-cubic, $4\times4\times4$ supercell at the $\Gamma$ point at 8 different temperatures (\SI{100}{K} -- \SI{800}{K}). 
    Each MD comprised 8000 snapshots (time step: \SI{8}{fs}) where we discarded the first 4000 in every MD and randomly selected 100 structures from the remaining 4000 for further evaluations. We have tested different sample sizes and equidistant sampling to ensure that our results for 100 snapshots are statistically converged.
    For each of these 100 snapshots, an additional DFT calculation was performed using a $3\times3\times3$ \vec{k}-point grid. 
    This was followed by non self-consistent calculations of the electronic bandstructure, which then was unfolded using the \texttt{BandUp} code \cite{medeiros_effects_2014, medeiros_unfolding_2015}.
    ML-MD calculations were performed in \texttt{VASP} using \SI{8}{\text{\r{A}}} as a cut-off radius for the radial descriptor and an increased weight of $10$ for force scaling. 
    All other settings were kept the same as in the DFT-MD. 
    
\section{Training of the TB model}\label{app:tbfit}
    Optimal TB parameters were found using the gradient-descent algorithm in combination with the \texttt{Adam} optimizer \cite{kingma_adam_2017} as implemented in Ref.~\citenum{innes_fashionable_2018}. 
    Gradients of errors with respect to TB parameters were calculated from finite-differences. Initial values of the TB parameters were set to 1 and the learning rate to $0.1$ and we set a cut-off radius of \SI{7}{\AA} for interactions. This value was set as large as possible, such that a larger cut-off produces the same Hamiltonian up to a tolerance of \SI{1e-5}{eV}.
    Mismatches between TB and DFT eigenvalues were calculated using the MAE with the option to assign a weight to each \vec{k}-point and band index. 
    To enhance the accuracy for the most relevant electronic-structure properties (i.e., all valence and lowest conduction bands), we decreased the weight of the upper two conduction bands to $0.25$ to decrease their influence on the overall fit. 
    Likewise, because GaAs is a direct bandgap semiconductor we chose to achieve higher accuracy for the bandgap than for the rest of the bandstructure. For this reason we increased the weighting of the bandgap value by increasing the weight of the $\Gamma$ point to $6$.
    The maximum number of iterations in this optimization was set to 2000.
    When comparing the TB bandgap data to experiments at various temperatures, we found that the effect of thermal expansion on the onsite parameters could not be neglected above ${\approx}$\SI{300}{K}.
    Therefore, we fitted the TB parameters again for each lattice constant to account for thermal expansion, which is particularly relevant for the onsite parameters which otherwise did not depend on atomic positions.
    In this comparison, when fitting the model to HSE, we reduced the weight of the upper two conduction bands to 0.1 and increased the weight of the upper valence band and lower conduction band to 2 to achieve higher accuracy for bandgaps.\\
    
\section{Effective bandstructure of disordered supercells}\label{app:effbands}
    We obtained effective bandstructures using band unfolding for bandstructures of each MD snapshot averaging over 100 snapshots at each temperature. In the Brillouin zone of the primitive cell, bands appear as peaks with finite width in the spectral function
    \begin{equation}\label{eq:specfunc}
         A(\vec{k}, E) = \sum_J W_{\vec{K}J}(\vec{G})\delta(E - E_{\vec{K}J}),
    \end{equation}
    where $W_{\vec{K}J}$ is the spectral weight of the $J$-th eigenvalue at $\vec{K}$ and $\vec{G}$ is the reciprocal lattice vector that unfolds $\vec{K}$ into $\vec{k}$. 
    Realizing an approach that was inspired by Refs.~\citenum{boykin_practical_2005, boykin_approximate_2007}, we implemented the unfolding method following Refs.~\citenum{lee_unfolding_2013} and \citenum{farjam_projection_2015}.
    Specifically, we used Eq.~17 from Ref.~\citenum{farjam_projection_2015} for computing spectral weights. In a final step, the effective bandstructure is obtained by identifying the positions of peaks and their width in the averaged spectral function. 
    We calculated the mismatch between bandgaps/effective bandstructures obtained from TB and DFT by using the MAE, i.e.,
    \begin{equation}\label{eq:mae}
         L\subtext{MAE} = \frac{1}{N_\varepsilon N_{\vec{k}}} \sum_{m\vec{k}} \abs{E^{\mathrm{DFT}}_m(\vec{k}) - E^\mathrm{TB}_m(\vec{k})},
    \end{equation}
    where $N_\varepsilon$ and $N_{\vec{k}}$ is the total number of energy eigenvalues and \vec{k}-points, respectively.
    For calculating converged temperature-dependent effective masses a small $\vec{k}$-point interval would have been required. 
    However, this was determined not to be meaningful because thermal energy fluctuations, in combination with remaining numerical inaccuracies of the TB model itself, were found to be of similar magnitude as the eigenvalues. 
    This implied large fluctuations of the effective masses and thermal trends that were non-meaningful.
    Hence, we chose using a larger $\vec{k}$-point interval when computing effective masses.
    Specifically, the effective mass for one MD snapshot was estimated fitting a parabola through 4 (3 for the light-hole) $\vec{k}$-points in direction $\Gamma$--$X$ of the unfolded Brillouin zone, using a $\vec{k}$-point step size of \SI{0.12}{\angstrom^{-1}}. 
    This larger step size led to an increase of the effective mass. 
    The reported effective masses were then obtained by averaging the data over 100 MD snapshots. 
    
\bibliographystyle{apsrev4-2}
\bibliography{literature.bib}
\end{document}